# Transport in disordered monolayer MoS$_2$ nanoflakes – evidence for inhomogeneous charge transport

Shun-Tsung Lo[1,2], O. Klochan[1], C.-H. Liu[3], W.-H. Wang[3], A. R. Hamilton[1,4], and C.-T. Liang[2,4]

[1]School of Physics, University of New South Wales, Sydney, NSW 2052, Australia
[2]Graduate Institute of Applied Physics, National Taiwan University, Taipei 106, Taiwan
[3]Institute of Atomic and Molecular Sciences, Academia Sinica, Taipei 106, Taiwan

**Abstract**
We study charge transport in a monolayer MoS$_2$ nanoflake over a wide range of carrier density, temperature, and electric bias. We find that the transport is best described by a percolating picture in which the disorder breaks translational invariance, breaking the system up into a series of puddles, rather than previous pictures in which the disorder is treated as homogeneous and uniform. Our work provides insight to a unified picture of charge transport in monolayer MoS$_2$ nanoflakes and contributes to the development of next-generation MoS$_2$–based devices.

Supplementary data files are available

[4]Correspondence to alex.hamilton@unsw.edu.au, ctliang@phys.ntu.edu.tw.

## 1. Introduction

Exploring a facile and controllable route to open a bandgap in graphene has been at the center of research in developing graphene-based nanoelectronics that is considered to be a promising candidate in the coming post-silicon era with continuing scaling down of device size to its physical limit. To date, a variety of methods such as chemical functionalization [1-6] and shape patterning [7-9] have been proposed to engineer the electrical properties of graphene. As an alternative option, layered transition metal dichalcogenide semiconductors such as molybdenum disulphide ($MoS_2$) have attracted renewed interest [10, 11]. In contrast to gapless graphene, two-dimensional (2D) monolayer $MoS_2$ has a direct bandgap of ~ 1.6-1.8 eV, which makes $MoS_2$-based field-effect transistors (FETs) attractive due to a high on-off current ratio [12, 13]. In addition, since monolayer $MoS_2$ shares a similar hexagonal structure and 2D nature but has different electrical properties from those of graphene, several fascinating applications such as high-gain phototransistor [14] and nonvolatile memory [15] were demonstrated based on $MoS_2$/graphene heterostructures. At present, a key challenge for $MoS_2$ to be applied in high-speed integrated circuits is the low carrier-mobility when deposited on substrates. Therefore much recent work has focused on improving the electrical performance of $MoS_2$ [16-18]. It has already been shown that encapsulation of $MoS_2$ in a high-κ dielectric environment, where charged-impurity scattering is effectively suppressed, leads to a significant increase in the carrier mobility [18]. To fully utilize $MoS_2$ in applications, fundamental studies of charge transport in monolayer $MoS_2$ are essential.

In atomically thin 2D systems such as graphene, studies using local probes indicate that at low carrier densities or when disorder is strong, the system can break up into charge puddles [19, 20]. Similar observations of the formation of charge puddles have also been reported in conventional semiconductor-based 2D systems [21, 22]. For a wide variety of 2D systems, when the carrier density is decreased, the system becomes inhomogeneous, giving rise to charge puddles and a percolation transition [23-25]. The carrier density at which the inhomogeneity sets in depends on the magnitude of disorder. In graphene, at a low carrier density close to the Dirac point, disorder originating from the randomly distributed trapped charges inside the substrate or near the graphene-substrate interface break up the 2D system into spatially inhomogeneous electron-hole puddles, even though they are not well isolated due to Klein tunnelling [26, 27]. If similar disorder were present in $MoS_2$ which has an intrinsic bandgap, it would be expected to break the system into inhomogeneous puddles of conducting electrons and insulating barriers.

Previous work on transport in MoS$_2$ has assumed the system to be uniform and shown that variable-range hopping (VRH) conduction is responsible for the observed behavior [28, 29]. On the other hand, by suppressing the Coulomb impurity potential, a crossover from insulating behavior where the measured conductance $G$ decreases with decreasing temperature $T$ ($\frac{dG}{dT} > 0$) to metallic behavior ($\frac{dG}{dT} < 0$) is clearly visible with increasing the Fermi energy $E_F$ [18]. The underlying mechanism responsible for transport on the insulating side of this transition was explained in terms of thermally activated behavior rather than VRH. Very recently, Qiu *et al*. [30] and Ghatak *et al*. [31] reported interesting results which demonstrate that structural inhomogeneity of MoS$_2$ itself also plays a role in the charge transport. Amongst these findings, a clear and unified picture of how disorder affects charge transport in MoS$_2$ is still lacking. In this paper, we perform detailed transport studies of an unencapsulated monolayer MoS$_2$ FET. We show that the experimental data can be well explained by the model of charge puddles. At high carrier densities and high temperatures, electron transport through charge puddles is the main mechanism. At very low temperatures or high disorder, transport *via* hopping occurs. Our work reconciles two seemly disparate views of inhomogeneous charge puddles and homogeneous hopping conduction in MoS$_2$ nanoflakes.

## 2. Experimental details

A monolayer MoS$_2$ nanoflake mechanically exfoliated from bulk MoS$_2$ crystal on the octadecyltrichlorosilane (OTS) functionalized SiO$_2$ (300 nm)/p$^+$-Si substrate [32, 33] was identified by Raman spectrum with a 532 nm excitation laser. To avoid including extra disorder during the fabrication process, we adopt a resist-free method to deposit Au contacts [34]. DC transport measurements were performed in a close-cycle cryostat using two source measure units (Keithley 2400) to provide the back-gate and source-drain biases. Before measurements, the device was annealed at 140 °C under vacuum for 24 h to remove moisture absorbed on the MoS$_2$ surface. At temperatures below 190K the current falls below the resolution of our instruments (using a Keithley 2400, the data is reliable for the measured current $I > 10^{-11}$ A and thus the conductance $G > 2 \times 10^{-10}$ S when $V_{SD} = 50$ mV is applied).

## 3. Results and discussion

We have performed transport measurements on a back-gated monolayer MoS$_2$ nanoflake with a room temperature mobility of ~0.6 cm$^2$/Vs.. The basic characterization of the device including the optical image and Raman spectrum is presented in the Supporting Information. Figure 1 shows the conductance $G$ as a function of back-gate voltage $V_{BG}$ at various $T$ from 300 K down to 90 K. The inset of Fig. 1 depicts the schematic diagram of the device. The conductance increases with increasing $V_{BG}$, showing typical n-type behavior. Figure 2(a) shows the corresponding conductance as a function of $1/T$ at different $V_{BG}$. We find that at high $V_{BG}$, the conductance increases with decreasing $T$, indicative of metallic behavior ($\frac{dG}{dT}<0$), which is followed by insulating behavior ($\frac{dG}{dT}>0$) below $T \sim 270$ K. In the insulating regime for 190 K $\leq T \leq$ 270 K, the conductance can be well described by thermally activated transport [18]

$$G(T) = G_b \exp(\frac{-E_b}{k_B T}), \quad (1)$$

where $E_b$ is the activation energy, $k_B$ is the Boltzmann constant, and $G_b$ is a prefactor. It is important to note that, in common with other studies, the data which can be well fitted by the activation model are also in good agreement with hopping since both of them predict an exponential $T$ dependence of $G$. Because of the limited temperature range of measurements, it is always difficult to distinguish between thermal activation and the different types of hopping. In general, hopping transport can be characterized as

$$G(T) = G_p \exp[-(\frac{T_p}{T})^p], \quad (2)$$

where $G_p$ is a prefactor, $T_p$ is a characteristic temperature, and $p$ is the exponent depending on the details of density of states (DOS) around $E_F$. For Mott VRH [35, 36], the DOS is constant, leading to $p = 1/3$ and $T_{1/3} = 13.8/(k_B N(E_F)\xi^2)$ in 2D whereas for Efros-Shklovskii (ES) VRH [37-39], which considers Coulomb interactions, the DOS vanishes linearly with energy around $E_F$ for a 2D system, leading to $p = 1/2$ and $T_{1/2} = 2.8e^2/(4\pi\varepsilon_r\varepsilon_0 k_B \xi)$. The vanishing DOS due to interaction is called the Coulomb gap $E_{CG}$. Here $N(E_F)$, $\xi$, e, $\varepsilon_r$, and $\varepsilon_0$ stand for the DOS at $E_F$, localization length, electronic charge, dielectric constant, and vacuum permittivity, respectively. In the following we address the issue of distinguishing thermal activation from VRH in our device, which helps us understand the influence of disorder on the 2D electron system in monolayer $MoS_2$.

It has been suggested that the conductivity prefactor $\sigma_p$ should be much smaller than $e^2/h$ in phonon-assisted VRH or is on the order of $e^2/h$ when electron-electron interactions are significant [40]. For example, in reduced graphene oxide sheets, the prefactor $\sigma_{1/2}$ for ES-VRH behavior is found to be around $2e^2/h$ [41]. From the intercept of the fitting lines in Figs. S3(a) and S3(b) together with the known length-to-width ratio ~ 1/6 in Fig. S1(a) in the Supporting Information, we can obtain $\sigma_{1/3}$ and $\sigma_{1/2}$ for Mott and ES VRH, respectively. The results at various $V_{BG}$ are presented in Fig. 2(b). We find that the obtained prefactor using the VRH model is much larger than $e^2/h$, inconsistent with the predictions of both phonon-assisted and interaction-assisted hopping described above. Instead, the prefactor determined from the fits to Eq. (1) is around $e^2/h$. In addition, in section 2 of the Supporting Information, we demonstrate that the weakening of $G(T)$ around $T = 190$ K is not due to the crossover between Mott and ES VRH. Since both the nearest neighbor hopping (NNH) and activated transport can give rise to an exponential $T$ dependence of the conductivity as shown in Fig. 2(a), we need to check which is the dominate transport mechanism for 190 K $\leq T \leq$ 270 K. It is well established that for disordered systems conduction can be separated into various temperature regimes, based on Mott's concept of a mobility edge [35, 36]:

i) At very high temperatures ($T$>280 K in our experiments) a large number of carriers around $E_F$ are excited to the delocalized states above the mobility edge $E_{<\mu>}$, we will observe metallic behavior due to electron-phonon interactions, which means that the conductance $G$ increases with decreasing $T$.

ii) Upon reducing the temperature the number of carriers that can be thermally activated to delocalized states decreases, giving conduction that follows an Arrhenius form $G(T) = G_a\exp(-E_a/kT)$. In the percolation model, the activation energy $E_a$ is equivalent to the potential barrier $E_b$ (= $E_{<\mu>}$ - $E_F$) which separates the charge puddles.

iii) Upon further decreasing the temperature these thermal excitation processes to the mobility edge are no longer possible, and conduction occurs by thermally assisted hopping between localized states. At first conduction occurs by hopping between adjacent localized states, the so-called NNH. Nearest neighbor hopping also gives $G(T) = G_h\exp(-E_h/kT)$.

In our experiments, the metallic behavior (mechanism (i)) is observed at high temperatures ($T > 280$ K). Therefore the thermally activated behaviour we observe in the interval 190 K $\leq T \leq$ 270 K cannot be due to NNH (mechanism (iii)), since activation of carriers from *localized* to *delocalized* states must come first (mechanism (ii)). Based on these experimental results, we can infer that activated transport, rather than VRH and NNH is the reasonable mechanism in our gated MoS$_2$ device for 190 K $\leq T \leq$ 270 K. More interestingly, thermal activation with a prefactor on the order of $e^2/h$ is usually observed in an inhomogeneous system considering transport through saddle point junction between charge puddles [42, 43]. This percolating picture has been demonstrated to be crucial in a variety of transition phenomena such as metal-insulator transition [23, 24], superconductor-insulator transition [44, 45], and quantum Hall plateau-plateau transition [46, 47].

The inset of Fig. 2(a) presents the activation energy $E_b$ as a function of $V_{BG}$, which is obtained from the slopes shown in Fig. 2(a). If the random potential is independent of carrier density, we would expect the effective barrier height to go as $<E_b> = <E_b^0 - E_F>$, where $E_b^0$ is the energy of percolation threshold. In two dimensions, $E_F$ is given by $n_s/N(E_F)$, where the carrier density $n_s$ is controlled by the back gate. Hence we expect $E_b$ to vary linearly with $V_{BG}$, as seen in the data. In monolayer MoS$_2$, the ideal value of density of states is $\frac{2m^*}{\pi\hbar^2}$ = 3.8 x 10$^{18}$ eV$^{-1}$m$^{-2}$. Using the parallel-plate capacitor model with $n = C_{ox}(\frac{V_{BG} - V_{BG}^t}{e})$ and assuming that $C_{ox}$ = 1.15 x 10$^{-4}$ F/m$^{-2}$ for the 300-nm-thick SiO$_2$ used, we can then estimate the ideal value of $\frac{dE_b}{dV_g}$ = 0.19 meV/V, which is smaller than the measured slope of 2.77 meV/V in the inset of Fig. 2(a). Here $m^*$, $\hbar$, $n$, $C_{ox}$, and $V_{BG}^t$ represent the effective mass (~ 0.45$m_0$ for MoS$_2$) [18], reduced Plank constant, carrier density, capacitance per unit area, and threshold voltage, respectively. The discrepancy between theory and experiment is ascribed to $E_b^0$ itself changing with $V_{BG}$, that is, the disorder potential is affected by the carrier density. On the other hand, in disordered MoS$_2$, charged disorder can reduce the carrier density, both by raising the bottom of the conduction band and by introducing lateral potential confinement, which creates charge puddles. Both of these can cause a much lower value of $N(E_F)$ and thereby a larger value of $\frac{dE_b}{dV_g}$ with respect to the ideal one. In section 3 of the Supporting Information, we further characterize the MoS$_2$ device by illumination and exposure to air.

We note that below 190 K the measured conductance is higher than that is given by Eq. (1), suggesting that there is an additional conduction mechanism at low $T$. After identifying the thermal activation as the dominant transport mechanism for 190 K ≤ $T$ ≤ 270 K, we also need to clarify the observed deviation from Eq. (1) for $T$ < 190 K. Since we already know $G_b$ and $E_b$ of Eq. (1) from the fits for 190 K ≤ $T$ ≤ 270 K, the net contribution of the second transport mechanism below $T$ = 190 K can be determined by subtracting $G_b \exp(\frac{-E_b}{k_B T})$ from the experimentally measured $G(T)$. In the Supporting Information, this second transport mechanism is strongly temperature-dependent. However it is hard to identify the exact nature of this second transport mechanism. Figures S6(a)-(c) show that $G - G_b \exp(\frac{-E_b}{k_B T})$ can be well described by not only $G_h \exp(\frac{-E_h}{k_B T})$ due to thermally assisted tunnelling but also VRH model according to Eq. (2). Figure S6(d) in the Supporting Information presents the obtained prefactors using the three different models, all of which are now reasonable. For clarity, we show that the red fitting curves in Fig. 2(a) demonstrate the good agreement of the conductance data with $G_b \exp(\frac{-E_b}{k_B T}) + G_h \exp(\frac{-E_h}{k_B T})$ below 190 K using the obtained parameters. Note that both ES and Mott VRH models can also provide satisfactory explanation for our data at $T < 190$ K as shown in the Supporting Information, Figs. S6(e) and S6(f), respectively. Therefore we cannot distinguish between different types of hopping in this regime. Nevertheless, our analysis clearly shows at high $T$ transport is by thermal activation over large barriers $E_b$, and at low $T$ tunnelling *via* impurities sets in.

To provide further evidence for the existence of charge puddles, we study the evolution of current-voltage characteristics $I(V_{SD})$ with temperature and back-gate bias in Fig. 3 and Fig. 4. The raw data at three representative temperatures $T = 300$ K, 200 K, and 100 K, are presented in the Supporting Information, Figs. S7(a)-(c). The $I(V_{SD})$ becomes nonlinear with decreasing $T$ at $V_{BG} = 50$ V. Since the measured $I$ does not show an exponential increase with $V_{SD}$, as seen from the semi-log plot of $I$ vs $V_{SD}$ in the Supporting Information, Fig. S8, we can rule out the possibility that the nonlinearity is due to the non-ideal ohmic contacts behaving as Schottky barrier [31]. Figure 3(a) then shows $I(V_{SD})$ at various $T$ from 300 K down to 20 K for $V_{BG} = 50$ V on a log-log scale. The solid lines are fits to power-law behavior with an exponent α, $I \propto V_{SD}^\alpha$, for 1 V $\leq V_{SD} \leq$ 2 V. At $T = 300$ K in the metallic regime, $\alpha = 1$ indicates that $I(V_{SD})$ remains ohmic even at high $V_{SD}$. Lowering $T$ to 200 K increases $\alpha$ up to 1.3, which is a signature of inhomogeneity of charge distribution [48]. Moreover, at around $T = 190$ K, deviation from thermal activation scheme occurs. If there are charge puddles, then at low $T$, where the thermal activation to mobility edge is not possible, then we should get percolation as seen in other 2D semiconductors. Middleton and Wingreen (MW) have considered percolation transport in a system comprised of arrays of metallic dots [49]. They find that the $I(V_{SD})$ characteristics obey a power-law scaling form

$$I \sim (\frac{V_{SD}}{V_t} - 1)^\zeta, \qquad (3)$$

with the threshold voltage $V_t$ and characteristic exponent $\zeta$ when $V_{SD} > V_t$. Below $V_t$, the current is suppressed. Experimentally, scaling behavior following Eq. (3) with a wide spread in $\zeta$ have been reported in a variety of inhomogeneous 2D systems such as metal dot arrays [50], metal nanocrystal arrays [51], graphene quantum dot arrays [52], and electrostatically defined quantum dot lattice in a GaAs-based 2D electron gas [53]. To apply MW model in analyzing the data, we need to first identify $V_t$. However, for $T > 60$ K, a clear threshold is absent in Fig. 3(a). We therefore choose $V_t$ such that the $I(V_{SD})$ traces of two successive temperatures can collapse on top of each other at high $V_{SD}$ by using Eq. (3). Figure 3(b) then shows a log-log plot of the current $I$ as a function of the scaling parameter $\frac{V_{SD}}{V_t} - 1$ at various $T$. We find that for 120 K $\leq T \leq 200$ K, at high $V_{SD}$ the $I(V_{SD})$ characteristics show power-law scaling-like behavior with $\zeta = 1.3$ [53], characteristics of a percolation transition, further evidence for existence of charge puddles in our system.

We note that $\zeta$ deviates from 1.3 when $T < 120$ K and increases with decreasing $T$. Figure 3(c) summarizes the determined $V_t$ as a function of $T$. Interestingly, it shows that $V_t$ increases dramatically when $T < 120$ K and approximately linearly with $T$ for $T \geq 120$ K. It has already been demonstrated both experimentally and theoretically that increasing background charge disorder can significantly enhance $V_t$ and invalidate the scaling law of Eq. (3) [51-55]. Figures 4(a) and 4(b) then show $I$ as a function of $V_{SD}$ and of the scaling parameter $\frac{V_{SD}}{V_t} - 1$ at various $V_{BG}$ for $T = 200$ K on a log-log scale. Similarly, $V_t$ is chosen such that the traces of two successive back-gate voltages with a step of 5 V can collapse on top of each other at high $V_{SD}$. The obtained $V_t$ as a function of $V_{BG}$ is presented in Fig. 4(c). We can also observe that $V_t$ increases rapidly when $V_{BG} < 35$ V and approximately linearly with $V_{BG}$ for $V_{BG} \geq 35$ V. Therefore we know that the effective disorder changes with $V_{BG}$ as inferred from the much larger slope of $\frac{dE_b}{dV_g}$ compared with the ideal value.

At high $V_{BG}$, it has been reported that the metallic behavior can sustain at much lower $T$ in high-mobility $MoS_2$ devices [18]. However, in an unencapsulated $MoS_2$ nanoflake, disorder arising from external charged impurities or structural inhomogeneity may cause spatial fluctuations in the conduction band edge $E_c$, breaking up the ideally uniform 2D electron system into a distribution of charge puddles. Our results suggest that we can develop a model of charge puddles in the monolayer $MoS_2$. The picture in Fig. 5 captures the essence of this model. The puddles locate at the regions of higher carrier density and are immersed into the background of trapped states. We note that the conduction paths which give lower resistance compared to the other ones in parallel will dominate the resulting transport behavior. As depicted in Fig. 5, at high temperatures and high densities, transport through puddles is relatively easy and thus predominantly governs $G(T)$. When the coupling of puddles is strong, there exist accessible extended states above the Mott mobility edge. Therefore the metallic properties of $MoS_2$ can result in the increase of $G$ with decreasing $T$ as found in Fig. 2(a). With decreasing $T$, thermal activation of carriers between puddles is gradually suppressed and the system shows insulating behavior described by Eq. (1). As demonstrated in Fig. 2(a), for $T < 190$ K, thermally assisted tunnelling or hopping conduction becomes more important as the temperature is lowered. We observe that $I(V_{SD})$ follows the scaling law of Eq. (3) with $\zeta = 1.3$ for 120 K $\leq T \leq$ 200 K, suggesting the importance of transport through puddles. At sufficiently low $T$ (< 120 K) when the carriers are difficult to conduct directly through puddles, thermally-assisted tunnelling between puddles mediated by the surrounding trapped states or hopping along the trapped states is preferred instead, which is depicted in Fig. 5. Since the carrier conduction is strongly hindered due to the existence of trapped sites, it is expected that the measured $I$ will decrease more rapidly with decreasing $V_{SD}$ than that predicted by Eq. (3) when hopping conduction is important. Hence, these trapped states cause an increase in $\zeta$ and eventually invalidate Eq. (3) with decreasing $T$ as shown in Fig. 3(b).

According to the MW model, the characteristic exponent $\zeta$ of the power-law scaling is determined by the effective dimensionality of the conducting channels. It is correct that MW model suggests that $\zeta = 1.6$ to 2 for 2D. However in 1D (i.e., only one preferred path or a small number of paths carriers a majority of the current), the MW model gives $\zeta = 1$. In our experiments $\zeta = 1.3$ was found. For example, in the presence of two different kinds of disorder (which create puddles and trapped states, respectively) as depicted in Fig. 5, it can be expected that 1D channel through puddles dominates the transport at low $T$, since multiple hops between the localized states are much less efficient than hopping between 2D puddles. The key point is that if the system becomes inhomogeneous the disorder breaks the translational symmetry – the system is no longer purely two-dimensional, and is between the 2D (1.6 < $\zeta$ < 2) and 1D ($\zeta = 1$) limits. The observed scaling behavior with $\zeta = 1.3$ in our measurements thereby indicates the inhomogeneity of the measured MoS$_2$ nanoflake. We note that the mobility of graphene prepared on an OTS functionalized SiO$_2$ substrate is only about two times higher than that of graphene prepared on a SiO$_2$ substrate [32]. Therefore although OTS functionalized SiO$_2$ can substantially reduce the disorder effect from the substrate, the substrate still affects the transport properties of 2D materials such as graphene and MoS$_2$. It is worth mentioning that the seminal work of Qiu *et al.* [30], which combine transport measurements, transmission electron microscopy, density functional theory, and tight-binding calculations, has demonstrated the importance of sulphur vacancies on the localization behavior in MoS$_2$. Accompanied by the presence of sulphur vacancies are the unsaturated electrons in the surrounding Mo atoms, which act as donors to make MoS$_2$ behave as an n-type semiconductor. Therefore we believe that both the substrate and intrinsic defects such as sulphur vacancies contribute to the formation of puddles in our MoS$_2$ devices.

As shown in the Supporting Information, Fig. S9, no clear Coulomb oscillations due to charging are observed at $T > 40$ K. Thus we can estimate the charging energy of the puddle $e^2/C$ to be less than the thermal energy $k_B T$ at 40 K where $k_B$ is the Boltzmann constant. Using the simple capacitor calculation $C = \frac{\varepsilon_o \varepsilon_r A}{d}$ where $A$ and $d$ are the puddle area and the thickness of the SiO$_2$ layer (=300 nm), respectively, this yields a length scale of $\approx 600$ nm. However, we wish to point out that within this approach, we only calculate the gate capacitance and the total capacitance has been under-estimated. Therefore this length scale represents the upper bound for the puddle size.

Either decreasing $T$ or $V_{BG}$ will reduce the conduction current of MoS$_2$. However, as shown in Fig. 4(b), the scaling behavior with the exponent 1.3 is shown to remain valid even down to $V_{BG} = 10$ V with $I < 1$ nA, which suggests that puddle physics may still govern the carrier transport at low $V_{BG}$. This is due to the fact that decreasing $T$ from 200 K to 20 K can change the ratio of $\frac{E_b}{k_B T}$ in Eq. (1) by a decade. However $\frac{E_b}{k_B T}$ would not be highly dependent on $V_{BG}$ since the variation of $E_b$ with $V_g$ is not huge (about 2 times when $V_{BG}$ is reduced from 50 V to 10 V) as in Fig. 2(c), which can explain the survival of the power-law scaling of Eq. (3) with $\zeta = 1.3$ at low $V_{BG}$ in Fig. 4(b). As another evidence for the inhomogeneity of MoS$_2$, the current exhibits weak peak structures in back-gate voltage when $T < 100$ K in the Supporting

Information, Fig. S9 [28, 46]. We therefore suggest that transport through puddles is dominant for $T > 120$ K and mediated by the surrounding trapped states below $T = 120$ K. Instead of all electronic states which show strongly localized behavior under the assumption of homogenous disorder, we demonstrate that charge puddles (inhomogeneity) can be present in a disordered monolayer $MoS_2$ nanoflake by studying the evolution of conduction current when varying measurement temperature, carrier density and source-drain bias. The combination effects of disorder from substrates and from structural inhomogeneity are expected to be the underlying reason. Our results may pave the way for obtaining a unified picture of carrier transport in monolayer $MoS_2$ nanoflakes, which is important for advanced applications of $MoS_2$ in transistors, biosensors, photodetectors, *etc*.

## 4. Conclusion

In conclusion, we have demonstrated that charge puddles can be present in a disordered monolayer $MoS_2$ nanoflake. Varying the carrier density and temperature effectively modifies the coupling between different puddles and thus changes the dominant transport mechanism correspondingly. At high carrier densities and temperatures, metallic behavior is observed since the correlation between puddles is strong. With decreasing temperature, thermal activation across the potential barrier which separates the puddles governs the transport. With further decreasing temperature, which decreases the correlation of puddles, deviation from simple activation scheme is found. We show that it is associated to thermally assisted tunnelling along trapped states, which provides an extra contribution to the transport at low temperatures. In the same temperature range, we find that the current-voltage characteristics tend to follow power-law scaling behavior with the exponent 1.3, which is a characteristic signature of percolation and further confirms the existence of charge puddles. These results contribute not only to our basic understanding of charge transport in monolayer $MoS_2$ but also to the development of next-generation nanoelectronics based on atomically thin two-dimensional materials.

**Acknowledgments**
We acknowledge funding from the Australian Research Council through the DP scheme. S.T.L. was funded by the National Science Council (NSC), Taiwan (grant number: NSC 102-2917-I-002-106). We would like to thank Dr. Jack Cochrane, Mr. Roy Li, and Mr. Fan-Hung Liu for experimental help. CTL acknowledges financial support from National Taiwan University (grant no. 103R890932).

**References**

[1] Elias D C, Nair R R, Mohiuddin T M G, Morozov S V, Blake P, Halsall M P, Ferrari A C, Boukhvalov D W, Katsnelson M I, Geim A K and Novoselov K S 2009 *Science* **323** 610

[2] Usachov D, Vilkov O, Grüneis A, Haberer D, Fedorov A, Adamchuk V K, Preobrajenski A B, Dudin P, Barinov A, Oehzelt M, Laubschat C and Vyalikh D V 2011 *Nano Lett.* **11** 5401

[3] Wei D, Liu Y, Wang Y, Zhang H, Huang L and Yu G 2009 *Nano Lett.* **9** 1752

[4] Eda G, Lin Y-Y, Mattevi C, Yamaguchi H, Chen H-A, Chen I S, Chen C-W and Chhowalla M 2010 *Adv. Mater.* **22** 505

[5] Lu Y-F, Lo S-T, Lin J-C, Zhang W, Lu J-Y, Liu F-H, Tseng C-M, Lee Y-H, Liang C-T and Li L-J 2013 *ACS Nano* **7** 6522

[6] Lo S-T, Chuang C, Puddy R K, Chen T M, Smith C G and Liang C T 2013 *Nanotechnology* **24** 165201

[7] Han M Y, Özyilmaz B, Zhang Y and Kim P 2007 *Phys. Rev. Lett.* **98** 206805

[8] Liang X, Jung Y-S, Wu S, Ismach A, Olynick D L, Cabrini S and Bokor J 2010 *Nano Lett.* **10** 2454

[9] Han M Y, Brant J C and Kim P 2010 *Phys. Rev. Lett.* **104** 056801

[10] Mak K F, Lee C, Hone J, Shan J and Heinz T F 2010 *Phys. Rev. Lett.* **105** 136805

[11] Radisavljevic B, Radenovic A, Brivio J, Giacometti V and Kis A 2011 *Nat. Nanotechnol.* **6** 147

[12] Radisavljevic B, Whitwick M B and Kis A 2011 *ACS Nano* **5** 9934

[13] Liu K-K, Zhang W, Lee Y-H, Lin Y-C, Chang M-T, Su C-Y, Chang C-S, Li H, Shi Y, Zhang H, Lai C-S and Li L-J 2012 *Nano Lett.* **12** 1538

[14] Roy K, Padmanabhan M, Goswami S, Sai T P, Ramalingam G, Raghavan S and Ghosh A 2013 *Nat. Nanotechnol.* **8** 826

[15] Bertolazzi S, Krasnozhon D and Kis A 2013 *ACS Nano* **7** 3246

[16] Kim S, Konar A, Hwang W-S, Lee J H, Lee J, Yang J, Jung C, Kim H, Yoo J-B, Choi J-Y, Jin Y W, Lee S Y, Jena D, Choi W and Kim K 2012 *Nat. Commun.* **3** 1011

[17] Yoon Y, Ganapathi K and Salahuddin S 2011 *Nano Lett.* **11** 3768

[18] Radisavljevic B and Kis A 2013 *Nat. Mater.* **12** 815

[19] Martin J, Akerman N, Ulbricht G, Lohmann T, Smet J H, von Klitzing K and Yacoby A 2008 *Nat. Phys.* **4** 144

[20] Zhang Y, Brar V W, Girit C, Zettl A and Crommie M F 2009 *Nat. Phys.* **5** 722

[21] Ilani S, Yacoby A, Mahalu D and Shtrikman H 2001 *Science* **292** 1354

[22] Allison G, Galaktionov E A, Savchenko A K, Safonov S S, Fogler M M,

Simmons M Y and Ritchie D A 2006 *Phys. Rev. Lett.* **96** 216407

[23] Shashkin A A, Dolgopolov V T, Kravchenko G V, Wendel M, Schuster R, Kotthaus J P, Haug R J, von Klitzing K, Ploog K, Nickel H and Schlapp W 1994 *Phys. Rev. Lett.* **73** 3141

[24] Tracy L A, Hwang E H, Eng K, Ten Eyck G A, Nordberg E P, Childs K, Carroll M S, Lilly M P and Das Sarma S 2009 *Phys. Rev. B* **79** 235307

[25] Martin J, Feldman B E, Weitz R T, Allen M T and Yacoby A 2010 *Phys. Rev. Lett.* **105** 256806

[26] Adam S, Hwang E H, Galitski V M and Das Sarma S 2007 *Proc. Natl. Acad. Sci. U.S.A.* **104** 18392

[27] Katsnelson M I, Novoselov K S and Geim A K 2006 *Nat. Phys.* **2** 620

[28] Ghatak S, Pal A N and Ghosh A 2011 *ACS Nano* **5** 7707

[29] Jariwala D, Sangwan V K, Late D J, Johns J E, Dravid V P, Marks T J, Lauhon L J and Hersam M C 2013 *Appl. Phys. Lett.* **102** 173107

[30] Qiu H, Xu T, Wang Z, Ren W, Nan H, Ni Z, Chen Q, Yuan S, Miao F, Song F, Long G, Shi Y, Sun L, Wang J and Wang X 2013 *Nat. Commun.* **4** 2642

[31] Ghatak S and Ghosh A 2013 *Appl. Phys. Lett.* **103** 122103

[32] Chen S-Y, Ho P-H, Shiue R-J, Chen C-W and Wang W-H 2012 *Nano Lett.* **12** 964

[33] Li Y, Xu C-Y, Hu P and Zhen L 2013 *ACS Nano* **7** 7795

[34] Liao L, Lin Y-C, Bao M, Cheng R, Bai J, Liu Y, Qu Y, Wang K L, Huang Y and Duan X 2010 *Nature* **467** 305

[35] Mott N F 1968 *J. Non-Cryst. Solids* **1** 1

[36] Mott N, Pepper M, Pollitt S, Wallis R H and Adkins C J 1975 *Proc. R. Soc. Lond. A* **345** 169

[37] Efros A L and Shklovskii B I 1975 *J. Phys. C* **8** L49

[38] Shklovskii B I and Efros A L 1984 *Electronic Properties of Doped Semiconductors* (Berlin: Springer-Verlag)

[39] Khondaker S I, Shlimak I S, Nicholls J T, Pepper M and Ritchie D A 1999 *Solid State Commun.* **109** 751

[40] Khondaker S I, Shlimak I S, Nicholls J T, Pepper M and Ritchie D A 1999 *Phys. Rev. B* **59** 4580

[41] Joung D and Khondaker S I 2012 *Phys. Rev. B* **86** 235423

[42] Shimshoni E, Auerbach A and Kapitulnik A 1998 *Phys. Rev. Lett.* **80** 3352

[43] Noh J P, Shimogishi F, Idutsu Y and Otsuka N 2004 *Phys. Rev. B* **69** 045321

[44] Dubi Y, Meir Y and Avishai Y 2007 *Nature* **449** 876

[45] Sacépé B, Chapelier C, Baturina T I, Vinokur V M, Baklanov M R and Sanquer M 2008 *Phys. Rev. Lett.* **101** 157006

[46] Cobden D H, Barnes C H W and Ford C J B 1999 *Phys. Rev. Lett.* **82** 4695

[47] Ilani S, Martin J, Teitelbaum E, Smet J H, Mahalu D, Umansky V and Yacoby A 2004 *Nature* **427** 328

[48] Meir Y 1999 *Phys. Rev. Lett.* **83** 3506

[49] Middleton A A and Wingreen N S 1993 *Phys. Rev. Lett.* **71** 3198

[50] Reichhardt C and Olson Reichhardt C J 2003 *Phys. Rev. Lett.* **90** 046802

[51] Parthasarathy R, Lin X-M and Jaeger H M 2001 *Phys. Rev. Lett.* **87** 186807

[52] Joung D, Zhai L and Khondaker S I 2011 *Phys. Rev. B* **83** 115323

[53] Goswami S, Aamir M A, Siegert C, Pepper M, Farrer I, Ritchie D A and Ghosh A 2012 *Phys. Rev. B* **85** 075427

[54] Parthasarathy R, Lin X-M, Elteto K, Rosenbaum T F and Jaeger H M 2004 *Phys. Rev. Lett.* **92** 076801

[55] Müller H-O, Katayama K and Mizuta H 1998 *J. Appl. Phys.* **84** 5603

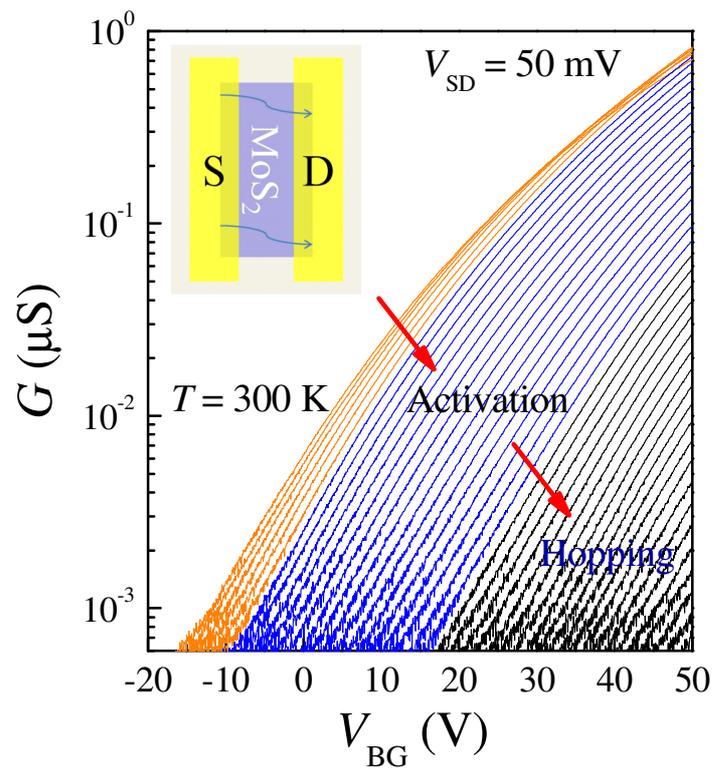

Figure 1 Conductance $G$ as a function of back-gate voltage $V_{BG}$ at various temperatures $T$ with a step of 5 K from 300 to 90 K. The arrow indicates the direction of decreasing $T$. Inset: schematic diagram of the device.

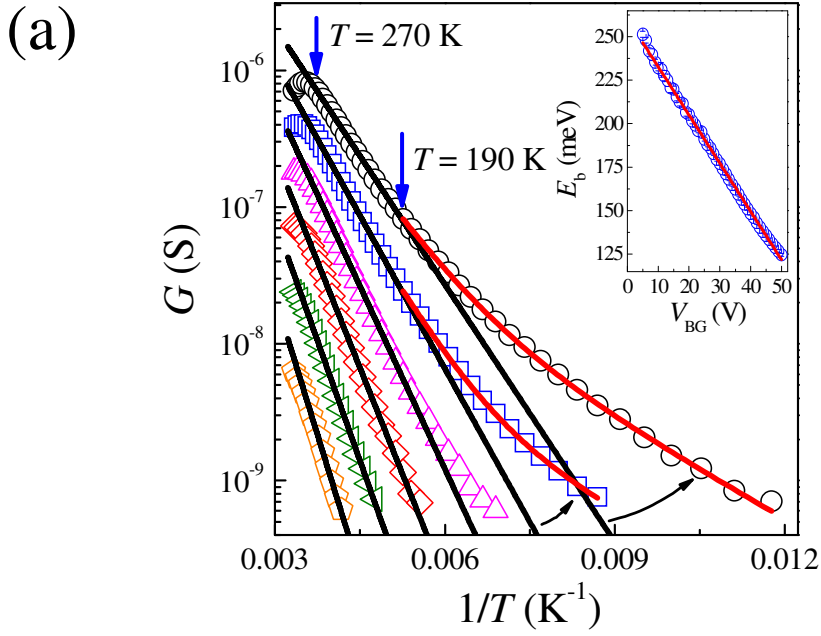

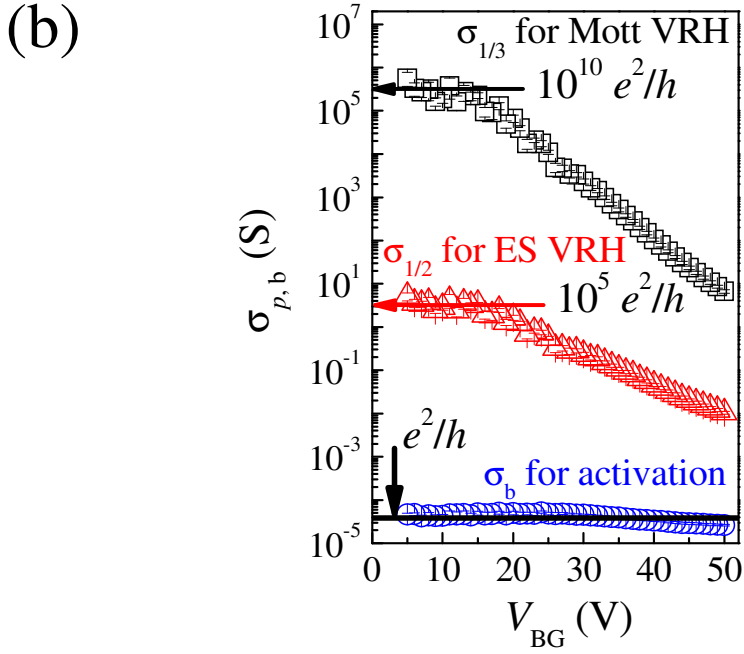

Figure 2 (a) Semi-log plot of $G$ versus $1/T$ at various $V_{BG}$. From top to bottom: $V_{BG}$ =50 V, 40 V, 30 V, 20 V, 10 V and 0. The black curves are the fits to Eq. (1) for 190 K $\leq T \leq$ 270 K. The red curves correspond to $G_b \exp(\frac{-E_b}{k_B T}) + G_h \exp(\frac{-E_h}{k_B T})$ for $T <$ 190 K using the parameters obtained from the fits to the data in the Supporting Information, Fig. S6(a). Inset: the obtained $E_b$ as a function of $V_{BG}$. (b) Prefactors $\sigma_p$ and $\sigma_b$ versus back-gate voltage $V_{BG}$, which is obtained from the fits in the Supporting Information, Fig. S3(a), Fig. S3(b), and Fig. 2(a) of the main text for Mott VRH ($\sigma_{1/3}$), ES VRH ($\sigma_{1/2}$), and activation ($\sigma_b$), respectively.

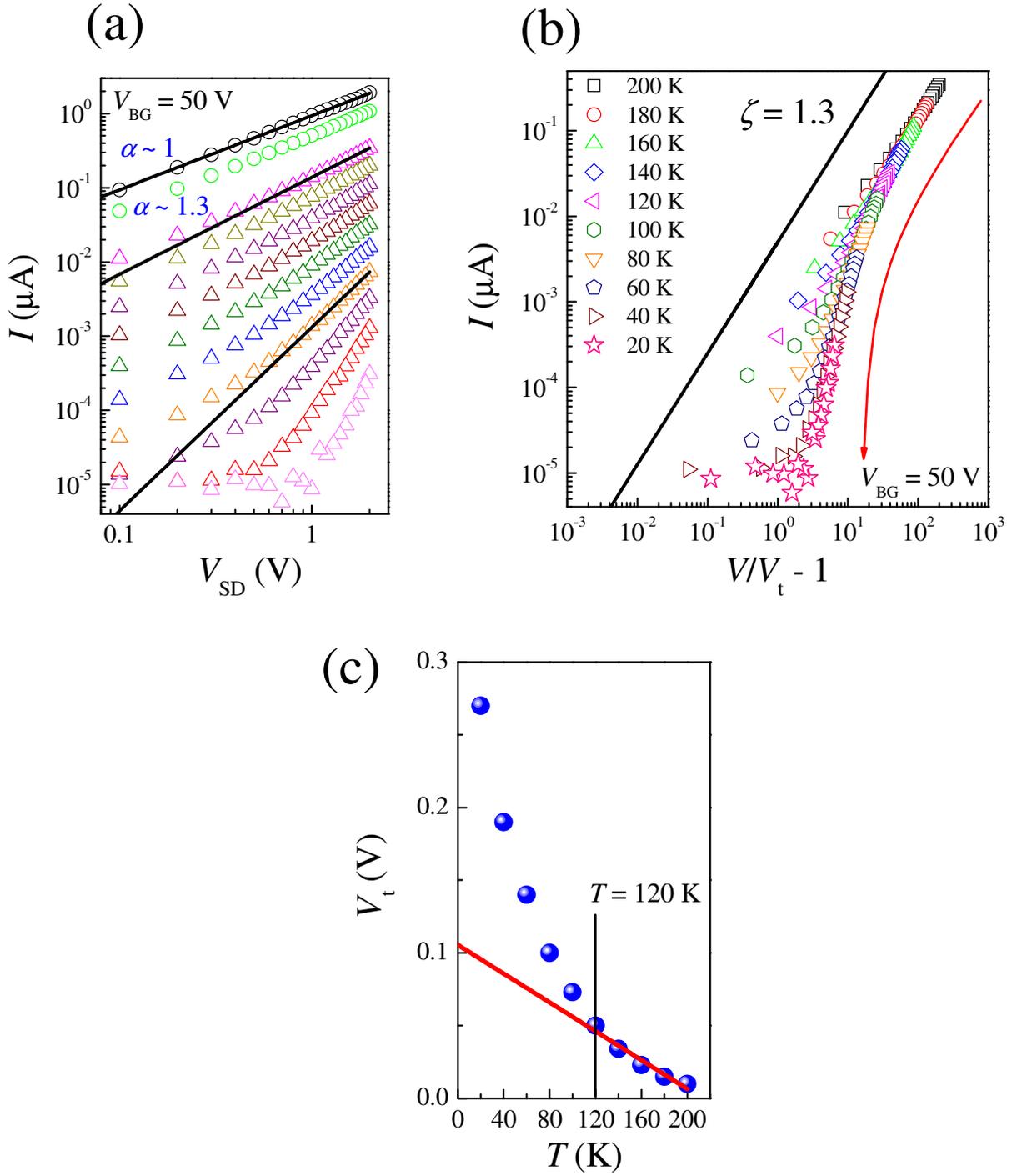

Figure 3 (a) Current-voltage characteristics $I(V_{SD})$ at various $T$, ranging from 300 K to 20 K for $V_{BG}$ = 50 V. From top to bottom: T=300 K, 250 K, 200 K, 180 K, 160 K, 140 K, 120 K, 100 K, 80 K, 60 K, 40 K and 20 K. Lines serve as a guide to the eye. (b) Log-log plot of $I$ versus $\frac{V_{SD}}{V_t} - 1$ for $T \leq 200$ K at $V_{BG}$ = 50 V. The solid line shows $\zeta$ = 1.3. (c) The corresponding $V_t$ as a function of $T$. The red line is a linear fit to the data for $T \geq 120$ K.

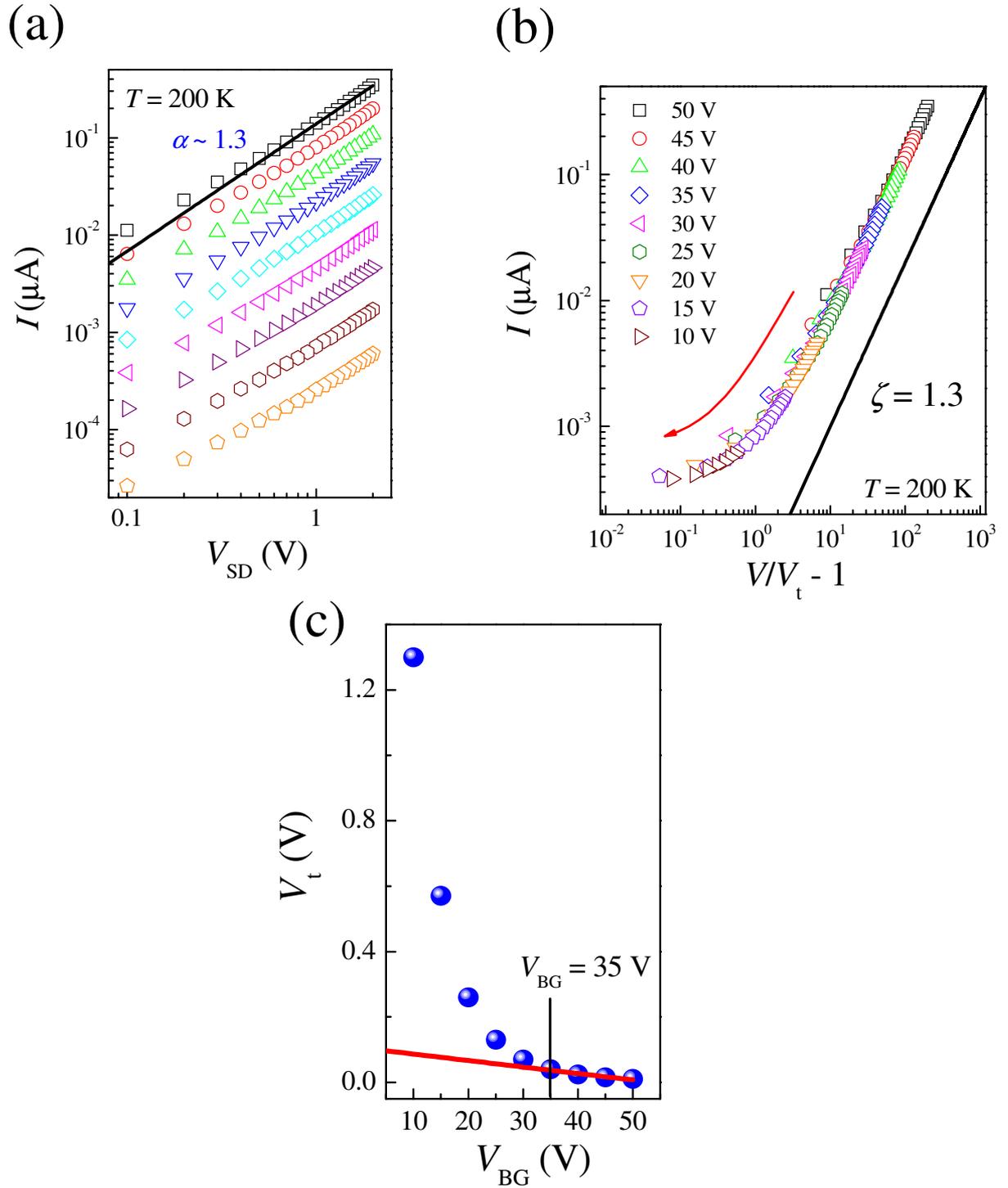

Figure 4 (a) Current-voltage characteristics $I(V_{SD})$ at various $V_{BG}$, ranging from 50 V to 10 V for $T = 200$ K. From top to bottom: $V_{BG}$ = 50 V, 45 V, 40 V, 35 V, 30 V, 25 V, 20 V, 15 V, and 10 V. (b) Log-log plot of $I$ versus $\frac{V_{SD}}{V_t} - 1$ at various $V_{BG}$, ranging from 50 V to 10 V for $T = 200$ K. The solid line shows $\zeta = 1.3$. (b) The corresponding $V_t$ as a function of $V_{BG}$. For $V_{BG} \geq 35$ V, the linear fit to the data is shown as the red line.

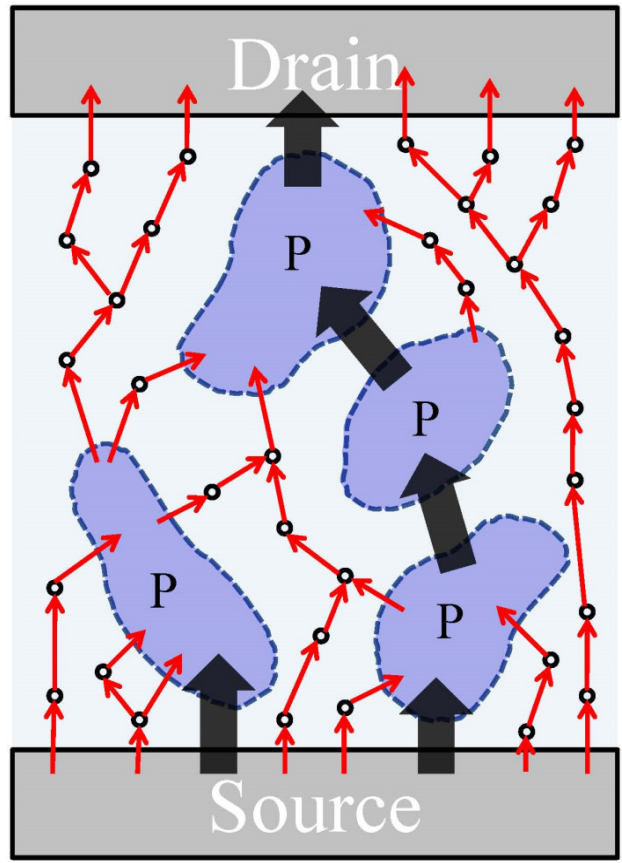

Figure 5 Schematic diagram illustrating different paths of carrier conduction. The shaded regions represent the charge puddles and the circles are for the trapped states. The black and red arrows correspond to the trajectories of transport through puddles and those of hopping process along trapped states, respectively.

# Transport in disordered monolayer MoS$_2$ nanoflakes – evidence for inhomogeneous charge transport
─ Supporting Information

Shun-Tsung Lo[1,2], O. Klochan[1], C.-H. Liu[3], W.-H. Wang[3], A. R. Hamilton[1,4] and C.-T. Liang[2,4]

[1]School of Physics, University of New South Wales, Sydney, NSW 2052, Australia
[2]Graduate Institute of Applied Physics, National Taiwan University, Taipei 106, Taiwan
[3]Institute of Atomic and Molecular Sciences, Academia Sinica, Taipei 106, Taiwan

[4]alex.hamilton@unsw.edu.au, ctliang@phys.ntu.edu.tw

## 1. Characterization of the measured device

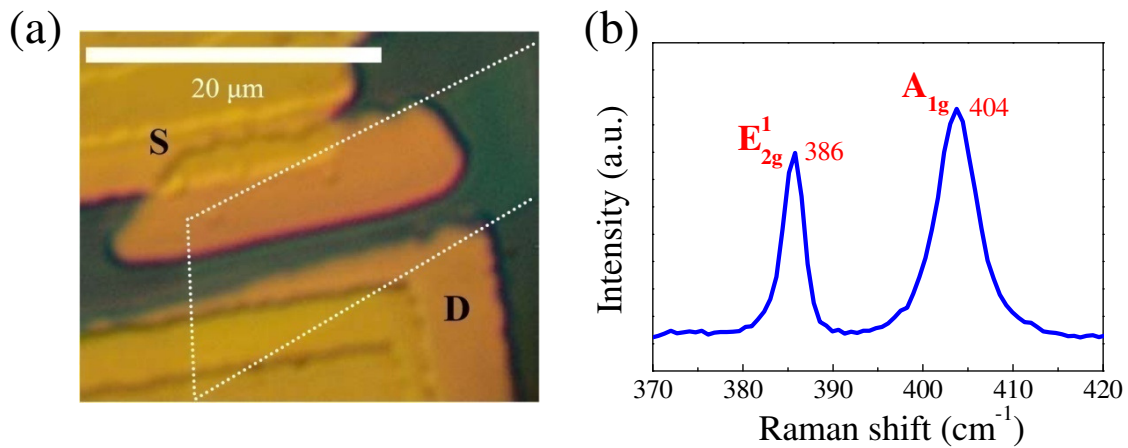

Figure S1 (a) Optical micrograph and (b) Raman spectrum of the measured MoS$_2$ device. The difference between two specific peaks assigned as $E^1_{2g}$ and $A_{1g}$ modes of Raman band is ~18 cm$^{-1}$, consistent with the reported value for monolayer MoS$_2$.

*Address correspondence to alex.hamilton@unsw.edu.au, ctliang@phys.ntu.edu.tw.

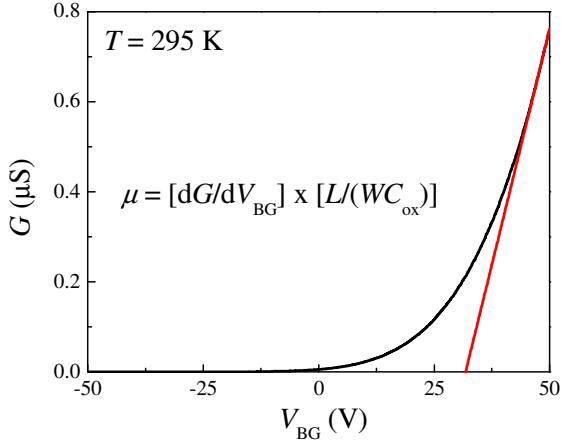

Figure S2 The room-temperature mobility of ~0.6 cm$^2$/Vs is extracted using the standard approach [1]. $\mu = [dG/dV_{BG}] \times [L/(WC_{ox})]$ with $L/W \sim 1/6$, $C_{ox} = 1.15 \times 10^{-4}$ F/m$^{-2}$, and $dG/dV_{BG}$ determined from the slope of the linear fit or 48 V $\leq V_{BG} \leq$ 50 V in Fig. S2.

## 2. Identification of the transport mechanism for 190 K < T < 270 K

Here we provide the second method to rule out variable-range hopping (VRH) as the dominant transport mechanism for 190 K < T < 270 K by considering the expected crossover behavior in the VRH regime. In general, in a gated device, the observation of ES VRH is limited at both high and low temperatures [2]. At high temperatures, carriers may have enough thermal energy to overcome $E_{CG}$, making the DOS effectively constant. Therefore, in this regime, Mott VRH will be observed. On the other hand, with decreasing temperature, the hopping length $R_h$ will continuously increase following $R_h \sim (\xi/4)(-(T_p/T)^p)$ [3]. At sufficiently low $T$, $R_h$ can be comparable to the thickness of oxide layer and thus Coulomb interaction will be screened by a metallic gate, which makes Mott VRH dominant again. Hence, in an ideal case, we can observe a crossover from Mott to ES VRH and then ES to Mott VRH with decreasing $T$. In our case, for 160 K $\leq T \leq$ 270 K both Mott and ES VRH can be well fitted to our results as shown in Figs. S3(a) and S3(b). Deviation from the fits is found at $T < 160$ K for $V_{BG} = 50$ V, indicating that there is a crossover of the transport mechanism with temperature. Since $T_{1/3}$ for Mott VRH is inversely proportional to $N(E_F)$, one can expect that the slope of ln$G$ versus $T^{-1/3}$ becomes larger in magnitude when the Coulomb gap, which diminishes $N(E_F)$, plays a role with decreasing $T$. Such a prediction has been verified in a delta-doped GaAs/Al$_x$Ga$_{1-x}$As heterostructure [3]. Our data, showing that the variation of $G$ weakens for $T < 160$ K in Fig. S3(a), are thereby inconsistent with this picture concerning the crossover from Mott to ES VRH transport. Moreover, a crossover from ES to Mott VRH with decreasing $T$ due to the screening effect should not occur since the hopping length (~

0.9 nm at $T = 90$ K for $V_{BG} = 50$ V), obtained from the fits with the help of ES VRH model in Fig. S3(b), is much shorter than the oxide thickness of 300 nm. The hopping length $R_h$ as a function of $T$ is presented in Fig. S3(c). Such results indicate that VRH scenario is not appropriate for 190 K $< T <$ 270 K.

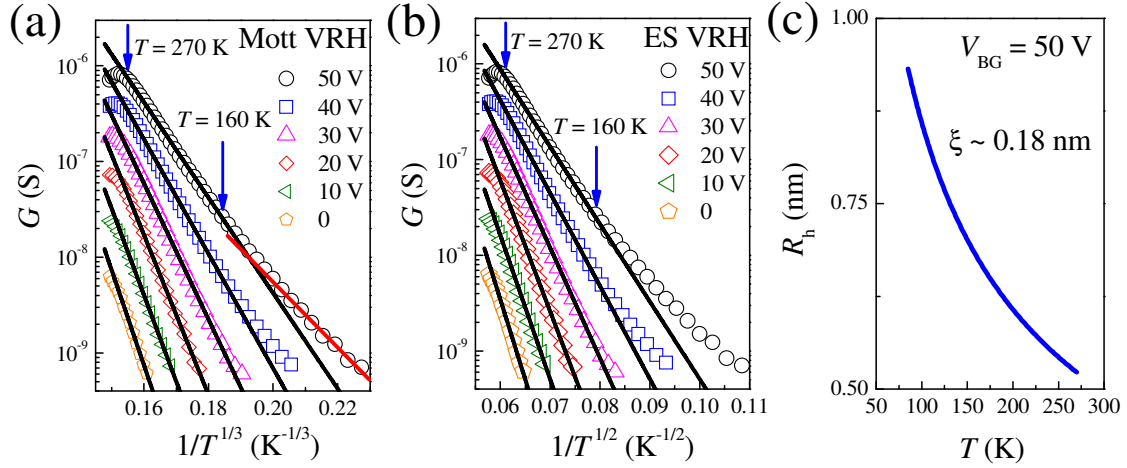

Figure S3 Semi-log plot of $G$ as a function of (a) $1/T^{1/3}$ and (b) $1/T^{1/2}$. (c) Hopping length $R_h$ derived from the fits in (b) as a function of $T$.

### 3. Additional data

To independently vary the activation energy $E_b$, we performed $T$-dependent measurements on the same device after illumination with a blue LED as well as after exposure to air. For $T < 210$ K the illumination caused a persistent change of the conductivity [4-6]. The data obtained after illumination and after exposure to air can be found in Fig. S5. The same fitting analyses were done and the results are presented in Fig. S5(c). As expected for $MoS_2$, illumination with blue light can excite the carriers from the valence band to the conductance band, which causes $n$-shift of $E_F$ and thereby reduces $E_b$. In contrast, $O_2$ and $H_2O$ molecules in air can deplete the electrons when they are physisorbed on the $MoS_2$ surface, which instead causes $p$-shift of $E_F$ and thereby increases $E_b$ as observed in Fig. S5(c) [7]. These results suggest that the experimentally-determined barrier height $E_b$ can be used to quantify the disorder within the $MoS_2$ system. As shown in Fig. S5(c), illumination or exposure process only moves $E_F$ without significantly changing the DOS.

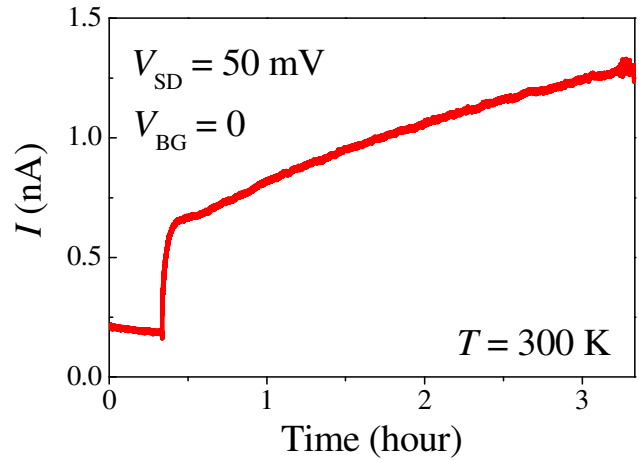

Figure S4 Photoresponse of the MoS$_2$ device under illumination of a commercial blue LED.

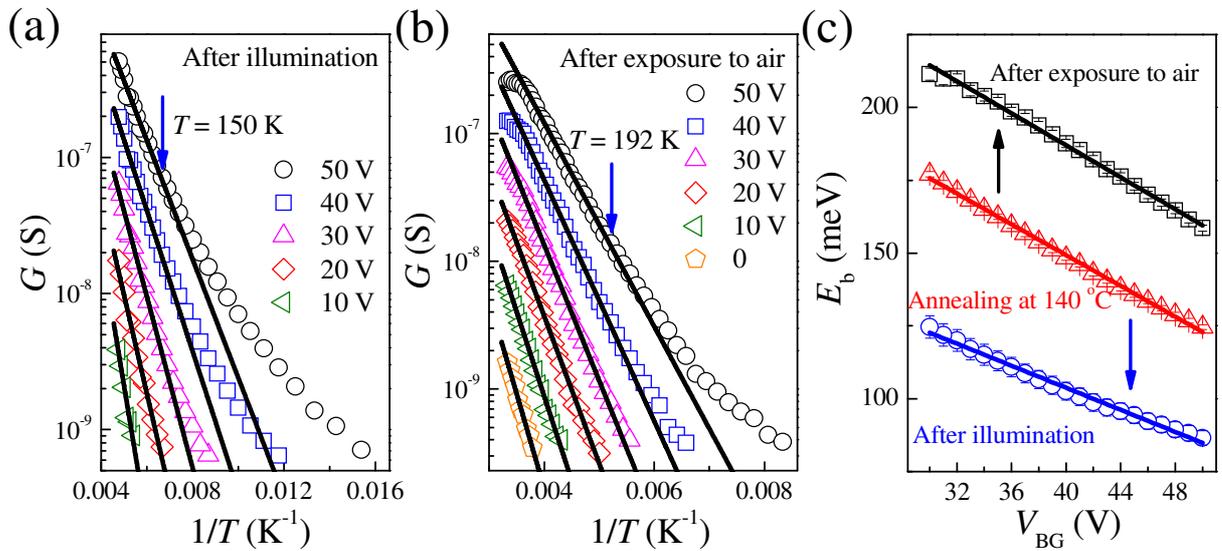

Figure S5 Semi-log plot of conductance $G$ versus $1/T$ obtained from the measurements on the same device (a) after the illumination process in Fig. S4 and (b) after exposure to air. The black curves are the fits to Eq. (1). (c) The comparison of the activation energy $E_b$ under different conditions.

## 4. Identification of the transport mechanism for $T < 190$ K

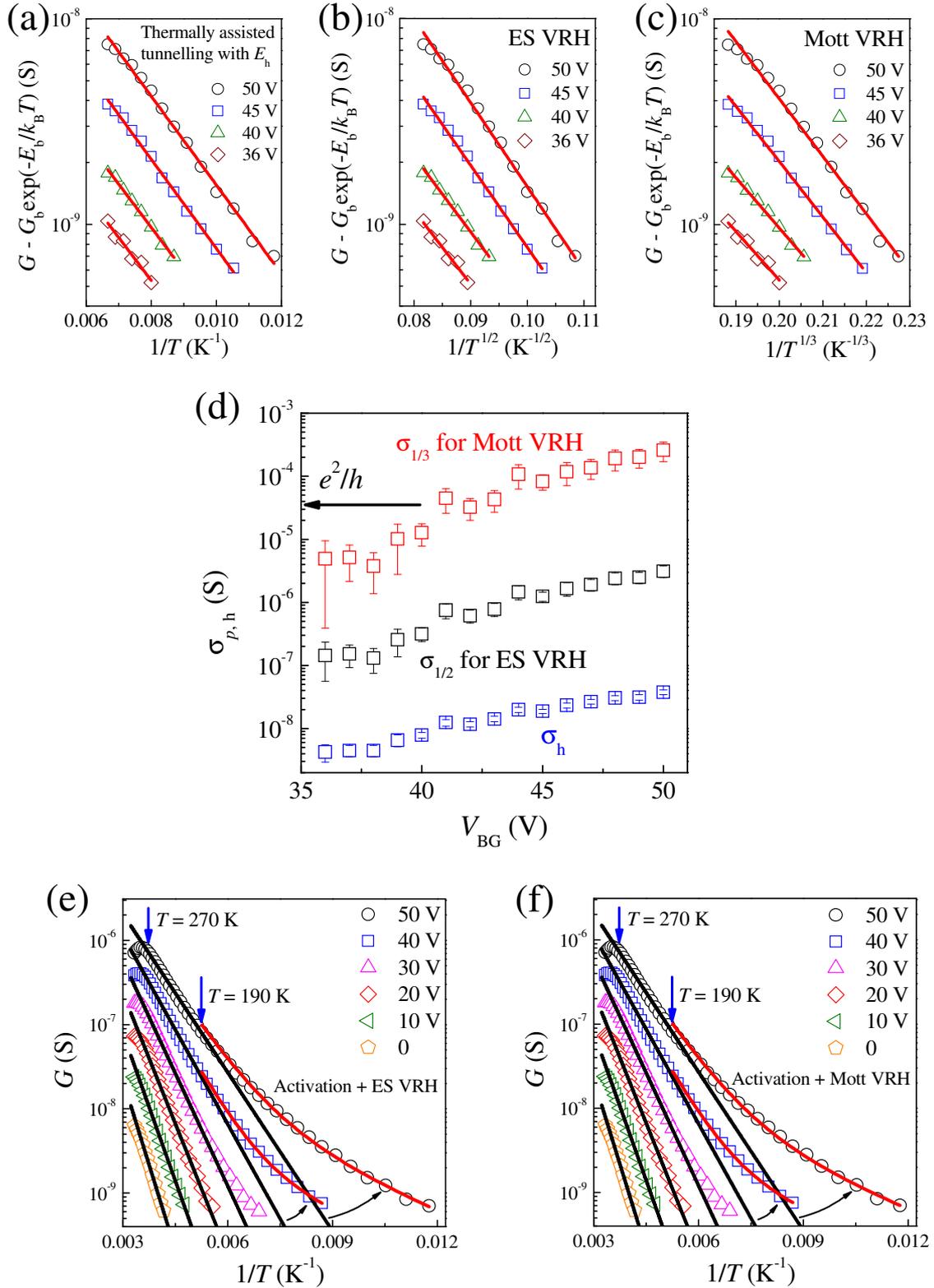

Figure S6 $G - G_b\exp(-E_b/kT)$ as a function of (a) $1/T$ (b) $1/T^{1/2}$ and (c) $1/T^{1/3}$ for 85 K $\leq T \leq$ 150 K. (d) Comparison of the obtained prefactors $\sigma_p$ and $\sigma_h$. (e), (f) The red curves correspond to $G_b\exp(-E_b/k_BT) + G_p\exp(-(T_p/T)^p)$ with $p = 1/2$ and $p = 1/3$ for ES and Mott VRH, respectively, and with the parameters determined in (b), (c), and (d).

## 5. Current-voltage characteristics

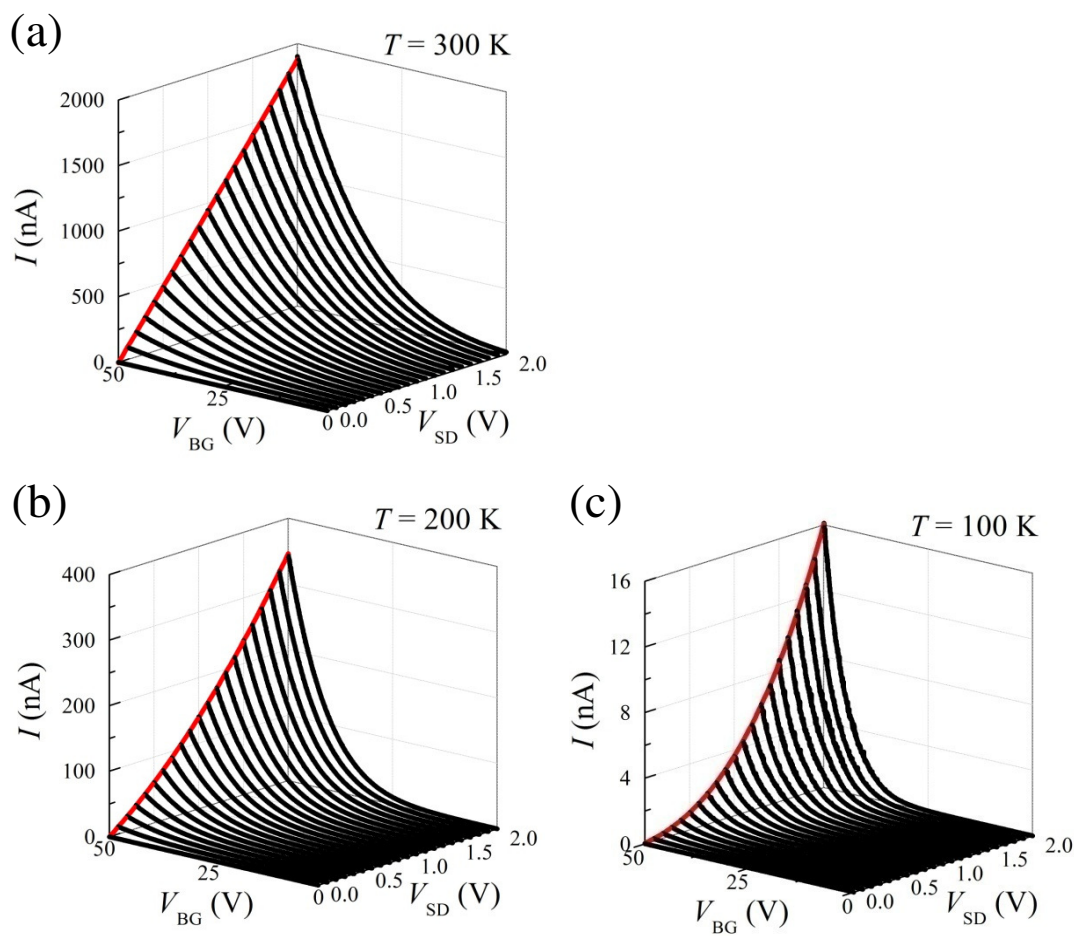

Figure S7 Current-voltage characteristics $I(V_{SD})$ at three representative temperatures of (a) $T = 300$ K, (b) 200 K, and (c) 100 K.

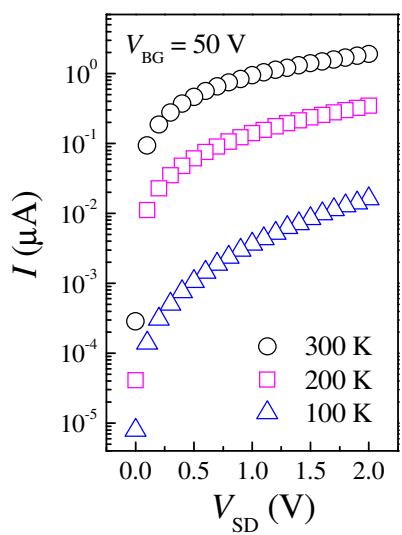

Figure S8 Semi-log plot of $I(V_{SD})$ at $T = 300$ K, 200 K, and 100 K for $V_{BG} = 50$ V.

## 6. Weak peak structure

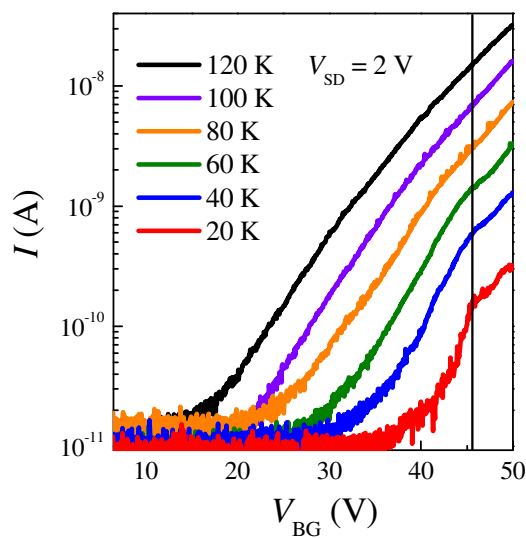

Figure S9 Semi-log plot of $I$ versus $V_{BG}$. The vertical line indicates the position of the weak peak structure.

## References

[1]  Radisavljevic B *et al.* 2011 Nat. Nanotech. **6**, 147; Fuhrer M S, Hone J 2013 Nature Nanotech. 8 146; Radisavljevic B, Kis A 2013 Nature Nanotech. **8** 147

[2]  Khondaker S I, Shlimak I S, Nicholls J T, Pepper M and Ritchie D A 1999 *Solid State Commun.* **109** 751

[3]  Khondaker S I, Shlimak I S, Nicholls J T, Pepper M and Ritchie D A 1999 *Phys. Rev. B* **59** 4580

[4]  Yin Z, Li H, Li H, Jiang L, Shi Y, Sun Y, Lu G, Zhang Q, Chen X and Zhang H 2011 *ACS Nano* **6** 74

[5]  Lopez-Sanchez O, Lembke D, Kayci M, Radenovic A and Kis A 2013 *Nat. Nanotechnol.* **8** 497

[6]  Wu C-C, Jariwala D, Sangwan V K, Marks T J, Hersam M C and Lauhon L J 2013 *J. Phys. Chem. Lett.* **4** 2508

[7]  Tongay S, Zhou J, Ataca C, Liu J, Kang J S, Matthews T S, You L, Li J, Grossman J C and Wu J 2013 *Nano Lett.* **13** 2831